\documentclass[AMA,STIX1COL]{WileyNJD-v2}

\articletype{Article Type}

\received{}
\revised{}
\accepted{}

\usepackage{microtype}
\usepackage{graphicx}
\usepackage{booktabs}
\usepackage[utf8]{inputenc}
\usepackage[T1]{fontenc}

\usepackage{amsmath,amsfonts,bm}

\def\eqref#1{equation~\ref{#1}}

\def\1{\bm{1}}

\DeclareMathAlphabet{\mathsfit}{\encodingdefault}{\sfdefault}{m}{sl}
\SetMathAlphabet{\mathsfit}{bold}{\encodingdefault}{\sfdefault}{bx}{n}

\newcommand{\E}{\mathbb{E}}

\usepackage{caption}
\usepackage{hyperref}
\usepackage{url}
\usepackage{graphicx}
\usepackage{float}
\usepackage{amsmath} 
\usepackage{algorithm}
\usepackage{multirow}
\usepackage{textpos}
\usepackage[font=small,labelfont=bf]{subcaption}

\newcommand{\liu}[1]{\textcolor{black}{#1}}

\makeatletter
\def\@fnsymbol#1{\ensuremath{\ifcase#1\or \dagger\or * \or \ddagger\or
   \mathsection\or \mathparagraph\or \|\or **\or \dagger\dagger
   \or \ddagger\ddagger \else\@ctrerr\fi}}
    \makeatother

\begin{document}

\title{~Large-scale Knowledge Distillation with Elastic Heterogeneous Computing Resources}

\author[1]{Ji Liu $^\dagger$}
\author[2]{Daxiang Dong $^\dagger$}
\author[3]{Xi Wang}
\author[4]{An Qin}
\author[5]{Xingjian Li}
\author[6]{Patrick Valduriez}
\author[7]{Dejing Dou}
\author[8]{Dianhai Yu}

\authormark{Liu \textsc{et al}}

\address[1]{\orgname{Baidu Inc.}, \orgaddress{\state{Beijing}, \country{China}}}
\address[6]{\orgname{Inria, University of Montpellier, CNRS, LIRMM}, \orgaddress{\state{Montpellier}, \country{France}}}

\corres{
$\dagger$ Equal contribution.\\
* Ji Liu. \email{liuji04@baidu.com}}

%
%

\abstract[Summary]{
Although more layers and more parameters generally improve the accuracy of the models, 
such big models generally have high computational complexity and require big memory, which exceed the capacity of small devices for inference and incurs long training time.
In addition, it is difficult to afford long training time and inference time of big models even in high performance servers, as well.
As an efficient approach to compress a large deep model (a teacher model) to a compact model (a student model), knowledge distillation emerges as a promising approach to deal with the big models. Existing knowledge distillation methods cannot exploit the elastic available computing resources and correspond to low efficiency.
In this paper, we propose an Elastic Deep Learning framework for knowledge Distillation, i.e., EDL-Dist. The advantages of EDL-Dist are three-fold. First, the inference and the training process is separated. Second, elastic available computing resources can be utilized to improve the efficiency. Third, fault-tolerance of the training and inference processes is supported.
We take extensive experimentation to show that the throughput of EDL-Dist is up to 3.125 times faster than the baseline method (online knowledge distillation) while the accuracy is similar or higher.
}


\maketitle

\begin{textblock*}{20cm}(3cm,6.8cm)
   {\large To appear in Concurrency and Computation: Practice and Experience}
\end{textblock*}

\section{Introduction}
\label{sec:intro}

Recent years have witnessed major success of deep Neural Networks (DNNs) in multiple domains, e.g., computer vision \cite{Villegas17, Szegedy2015}, natural language processing \cite{Devlin19,sun2021ernie}, and bio-informatics \cite{Caruana2015}.
More layers, neurons and parameters generally correspond to higher accuracy while making the model bigger. For instance, BERT \cite{Devlin19} and ERNIE \cite{sun2021ernie} exploit large numbers of parameters, e.g., from 110 million to 340 million parameters for BERT \cite{Devlin2018} and 10 billions parameters for ERNIE.
However, the big models generally have high computational complexity and require big memory, which exceed the capacity of small devices (mobile phones or IoT devices).
In addition, it is also hard to afford the long training time or the inference time of big models.

Introduced in \cite{Bucila2006} and generalized by \cite{Hinton2015}, as an efficient approach to distill the knowledge from a cumbersome model into a compact model \cite{Gou2020}, knowledge distillation retains the accuracy at the same time \cite{Mirzadeh2019,li2021knowledge}.
Knowledge distillation trains a small model (a student model) under the supervision of a big model (a teacher model). The teacher model is a cumbersome model, which could be an ensemble of separately trained models or a single very large model trained with a strong regularizer \cite{Hinton2015}. The student model is relatively small and compact, which correspond to short inference time and requires small storage space.

A normal deep neural network is trained during the training process, after which, it is ready for inference, e.g., classification or prediction. The process to train the student model is the training process while the process to generate supervision knowledge from the teacher model is the inference process.
Both the training process and the inference process exist within the knowledge distillation, while the training process of the student model is carried out using the results of the inference process of the teacher model. These inference and training processes can be performed sequentially or in parallel. Based on the training time of a teacher model, there are typically two methods to perform knowledge distillation \cite{Gou2020}. The first method is to train a student model with the supervision of a pre-trained teacher model. The second way is to train the teacher model and the student model in parallel when there is no available pre-trained teacher model. In this paper, we focus on the first method. 

The training process of the knowledge distillation can be typically carried out using two approaches, i.e., online and offline. The knowledge distilled from the teacher model is cached in an extra data store with the offline approach and is utilized to train the student model separately \cite{Gou2020}. Although it decouples the inference process and the training process, this approach requires much extra storage while it takes much time to distill the knowledge from the teacher model with big input data.
The teacher model and the student model are deployed in the same server with the online approach while the training process and the inference process are synchronously performed.
However, the training of the student model is restricted by the synchronization of the inference of the big teacher model.
Furthermore, these two approaches cannot exploit elastic computing resources to improve the efficiency and cannot support the fault-tolerance.

Diverse computing resources, e.g., CPU cores or GPU cards can be exploited for the training process of knowledge distillation. 
However, the computing resources may dynamically become unavailable because of other concurrent users.
Therefore, a user can utilize some dedicated computing resources for a long time while some elastic computing resources can only be ensured for a short time.
In addition the elastic computing resources may become unavailable because of other tasks of high priority. Furthermore, some computing resources may become unavailable due to exceptions, network connection problems or other unexpected issues, within the long training process of knowledge distillation. Some computing resources may also dynamically become available during the training process of the knowledge distillation. The dynamically varying computing resources cards are elastic resources. As diverse computing resources, e.g., CPUs and GPUs of diverse types, exist in a data center, the elastic resources are generally heterogeneous. \liu{Furthermore, it is critical to exploit numerous computing resources to enable large-scale knowledge distillation so as to achieve a short execution time. Large-scale knowledge distillation refers to the scale of distributed computing environment.} As a result, the problem of exploiting heterogeneous elastic computing resources, and ensuring the failures-tolerance of the system to perform \liu{large-scale} knowledge distillation becomes critical for the training process. 

In this paper, we address the problem of how to efficiently perform \liu{large-scale} knowledge distillation in a dynamically varying heterogeneous environment. 
We assume that two types of computing resources, i.e., elastic and dedicated, exist in a distributed environment, which can be utilized for the knowledge distillation. 
The dedicated computing resources are of high performance, e.g, V100\footnote{Parameters for V100: \url{https://www.nvidia.com/en-us/data-center/v100/}} GPU cards, for knowledge distillation only.
The elastic computing resources may correspond to relatively lower performance, e.g., P4\footnote{Parameters for P4: \url{https://images.nvidia.com/content/pdf/tesla/184457-Tesla-P4-Datasheet-NV-Final-Letter-Web.pdf}} GPU cards, and can be dynamically allocated to other tasks of higher priority or knowledge distillation.
In order to address the problem, we propose an Elastic Deep Learning framework, i.e., EDL-Dist.
We design a distributed architecture within EDL-Dist with the methods to address the fault-tolerant during the knowledge distillation.
We exploit an elastic service to manage multiple computing resources for the inference process of teacher models in EDL-Dist.
EDL-Dist manages the distributed training utilizing multiple GPU cards of diverse servers (a server may have one GPU or more GPU cards) during the training process of knowledge distillation. 
The training and inference are decoupled in order to improve the efficiency and exploit heterogeneous computing resources.
Inspired by the fail-over mechanism \cite{ozsu2020principles} we use check-points and re-execution of tasks to address the fault-tolerance.
We propose two main algorithms for data scheduling distributed training of knowledge distillation. 
While combining static \cite{liu2020two} and dynamic scheduling, the scheduling algorithm  associates computing resources from different processes.
The distributed training algorithm, which is denoted EDL-Dist, trains the student model in a decentralized way with the supervision of the distilled knowledge from the teacher model.
\liu{In addition, the combination of the decoupled architecture and the distributed training algorithm enables large-scale knowledge distillation with numerous computing resources.}
To the best of our knowledge, we are the first to enable distributed training for knowledge distillation with elastic heterogeneous computing resources.
We extend the conference version \cite{dong2022elastic} by adding more experimentation and discussions on the limitations and future work.
We summarize our contributions as follows:

\begin{figure}[htbp]
    \centering
    \includegraphics[width=0.8\textwidth]{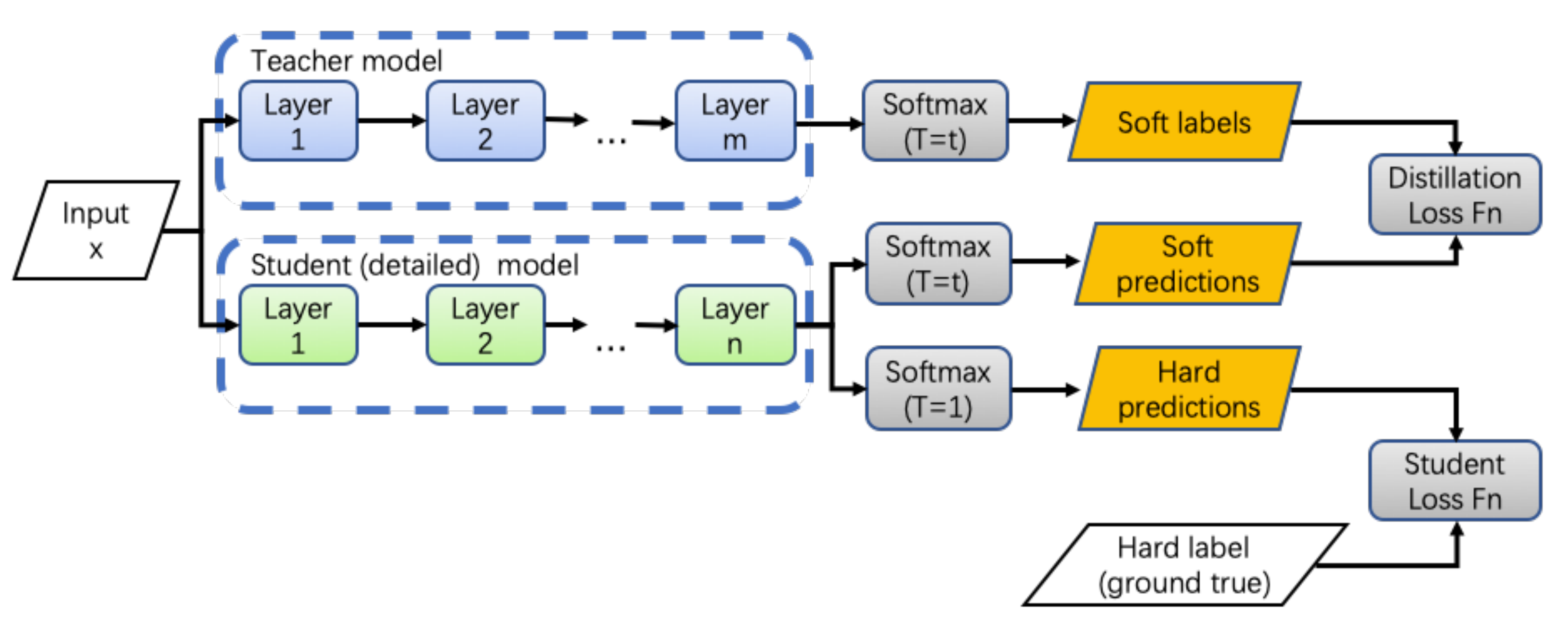}
    \caption{Knowledge distillation \cite{Zmora2019}.}
    \label{fig:knowledgedist}
\end{figure}

\begin{itemize}
	\item We propose the architecture of EDL-Dist, which consists of a student module, a teacher module and a coordinator module. This architecture decouples the training process in the student module and the inference process in the teacher module, which can accelerate the training process in the student module with elastic heterogeneous resources in the teacher modules.
	\item We propose dynamic hybrid resource scheduling method, the design of fault-tolerance and the EDL-Dist algorithm. The dynamic hybrid resource scheduling method can choose appropriate teacher GPU cards for each student GPU card while avoiding the risk to store many unused soft labels (see details in Section \ref{sec:background}). The design of fault-tolerance enables knowledge distillation with the consideration of the failures of teacher GPU cards during the training process. In addition, the design exploits a checkpoint mechanism to continue the training process when there are failures or exceptions in the student module. The EDL-Dist algorithm enables the training process with student GPU cards distributed in multiple servers while exploiting the soft labels from the teacher module.
	\item We carry out extensive experimentation (with up to 32 V100 and 614 K1200 GPU cards) to show the advantage of EDL-Dist. The experiment results show that the throughput of EDL-Dist can be similar to that of normal training and much bigger (up to 181\%) than that of online knowledge distillation. Within the normal training, the training process is performed with multiple GPU cards without the supervision of the knowledge from a teacher model. The training time of EDL-Dist with fine-tuned teacher GPU resources can be similar to that of normal training while it is much shorter than that of normal training. In addition, the accuracy of EDL-Dist is slightly higher than that of normal training as mentioned in \cite{Hinton2015}.
\end{itemize}

This paper is organized as follows. In Section \ref{sec:background}, we introduce the related work and background. In Section \ref{sec:edl}, we present the EDL-Dist framework. In Section \ref{sec:exp}, we show the experimental results, which demonstrate the advantage of EDL-Dist compared with the baseline method (the online approach) and normal training in terms of throughput and accuracy.
In Section \ref{sec:discussion}, we analysis the limitations of the current solution and present the future directions.
Finally, Section \ref{sec:con} concludes.

\section{Related Work}
\label{sec:background}

In this section, we present the background of elastic deep learning distillation, i.e., knowledge distillation.
Then, we introduce the methods for distributed and decentralized training and elastic computing resources.

\begin{figure*}[htbp]
    \centering
    \includegraphics[width=0.5\textwidth]{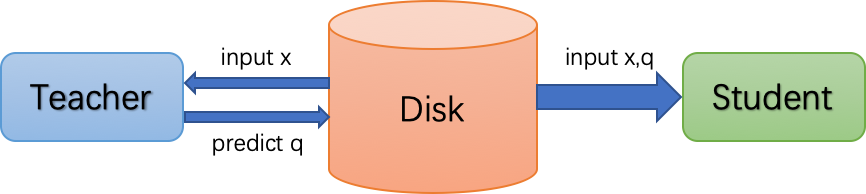}
    \caption{Offline knowledge distillation.}
    \label{fig:offline}
\end{figure*}

Model compression based on knowledge transferring was first proposed in \cite{Bucila2006} in order to compress an ensemble of models or a large model to a compact model.
The ensemble of models or the large model take much storage space and require a long time for inference while the compact model requires relatively small storage space and a short time for inference. 
Knowledge distillation is based on the popular machine learning Softmax function and a temperature \cite{Hinton2015}, which is defined in Formula \ref{eq:logit}.
\begin{equation}
\begin{split}
q_i = \dfrac{exp(z_i/T)}{\sum_{j}exp(z_j/T)}
\end{split}
\label{eq:logit}
\end{equation},
where $z_i$ is the output from the output layer of a teacher neural network; $T$ is a temperature, which indicates the impact of the output from the teacher model. 
The Softmax output layer computes the probability that the input data corresponds to each class with the corresponding computed logit.
The probability is related to a temperature \textit{T}, i.e., which represents the impact of the distilled knowledge of the teacher model on the student model. A higher temperature corresponds to a weaker impact.

As shown in Figure \ref{fig:knowledgedist}, during the training of knowledge distillation, two neural networks are used: teacher model and student model. 
The student model is trained based on the combination of two loss values. 
The first loss value is calculated from a soft prediction, which contains the probability of each class calculated from the teacher model. 
The soft prediction is calculated using Formula \ref{eq:logit}.
The other loss value corresponds to a hard prediction, which is the ground truth label from the training data. 

Two types of work are also related to this paper, i.e., mutual knowledge distillation and model pruning.
When there is no available teacher model, the teacher and student can be trained at the same time, i.e., an ensemble of students can learn collaboratively and teach each other throughout the training process \cite{Zhang2018,Chen2020}, which is the mutual knowledge distillation \cite{Gou2020}.
Note that the mutual knowledge distillation is also denoted ``online knowledge distillation'' in \cite{Gou2020}. 
However, we use the ``online knowledge distillation'' to represent the knowledge distillation with the teacher model and the student model deployed in the same GPU card during the training process of knowledge distillation in this paper.
Model pruning is also a practical method to reduce the size of a neural network while it requires manual configuration of sensitivity for layers, which is cumbersome \cite{Cheng2017}.

In order to accelerate the training process of a deep learning network, multiple GPUs can be exploited during the training process using data parallelism, where the model is replicated in each GPU while the data is distributed in different GPUs \cite{Anil2018}. Model parallelism \cite{Madiajagan2019} and pipeline parallelism \cite{Narayanan2019} exist for the training process with multiple GPUs while they are beyond the scope of this paper. There are basically two approaches to exploit multiple GPUs for the training process, i.e., distributed and decentralized. The distributed approach generally uses a centralized server to synchronize or schedule the execution of workers. Parameter server \cite{li2014communication,liu2021heterps} is a typical distributed approach. With this approach, the gradients or the weights of each worker is sent to a centralized server, which updates global gradients or weights and returns the updated global gradients or weights to each worker. 
There are diverse methods to synchronize the execution in each node, e.g., Bulk Synchronized Parallel (BSP) \cite{Valiant1990}, Asynchronous Parallel (ASP) \cite{Zhu2020} and Stale Synchronous Parallel (SSP) \cite{Ho2013}, in order to achieve load balancing among different computing nodes.
However, in this way, the centralized server becomes a single-point of failure and a bottle neck of the training. The decentralized approach does not reply on a centralized server while making each computing node equally perform the calculation based on a predefined protocol. The decentralized approach can be synchronized or asynchronized \cite{Lian2018}. The synchronized decentralized approach generally synchronizes within each iteration of the execution in all the workers based on an algorithm, e.g., Ring all-reduce \cite{ring-allreduce}. The asynchronized decentralized approach doees not synchronize can tolerate the late update of gradients or weights, which can avoid the synchronization within each iteration, e.g., D-PSGD \cite{Lian2017}. However, the implementations of Ring all-reduce, e.g. Horovod \cite{sergeev2018horovod}, and D-PSGD are only designed for the training process without the consideration of knowledge distillation. In addition, they cannot support elastic resources or fault-tolerance during the training process. 

Two existing methods, i.e., online and offline, can be exploited to carry out knowledge distillation. The offline method refers to training a teacher model before distillation, the knowledge of which can be extracted and the corresponding soft labels can be stored in a cache \cite{Gou2020}, as shown in Figure \ref{fig:offline}. However, extra storage space is need for this method. The online method combines the inference and the training process and deploys the implementation in the same GPU card in order to reduce the data transfer between the teacher module and student module \cite{Zmora2019}. In this situation, 
a big teacher model incurs long time to synchronize the training process and the inference, which reduces the efficiency. In addition, elastic computing resources cannot be exploited in these two methods and they cannot address the fault-tolerance issue during the training process of knowledge distillation.

\begin{figure*}[htbp]
    \centering
    \includegraphics[width=0.6\textwidth]{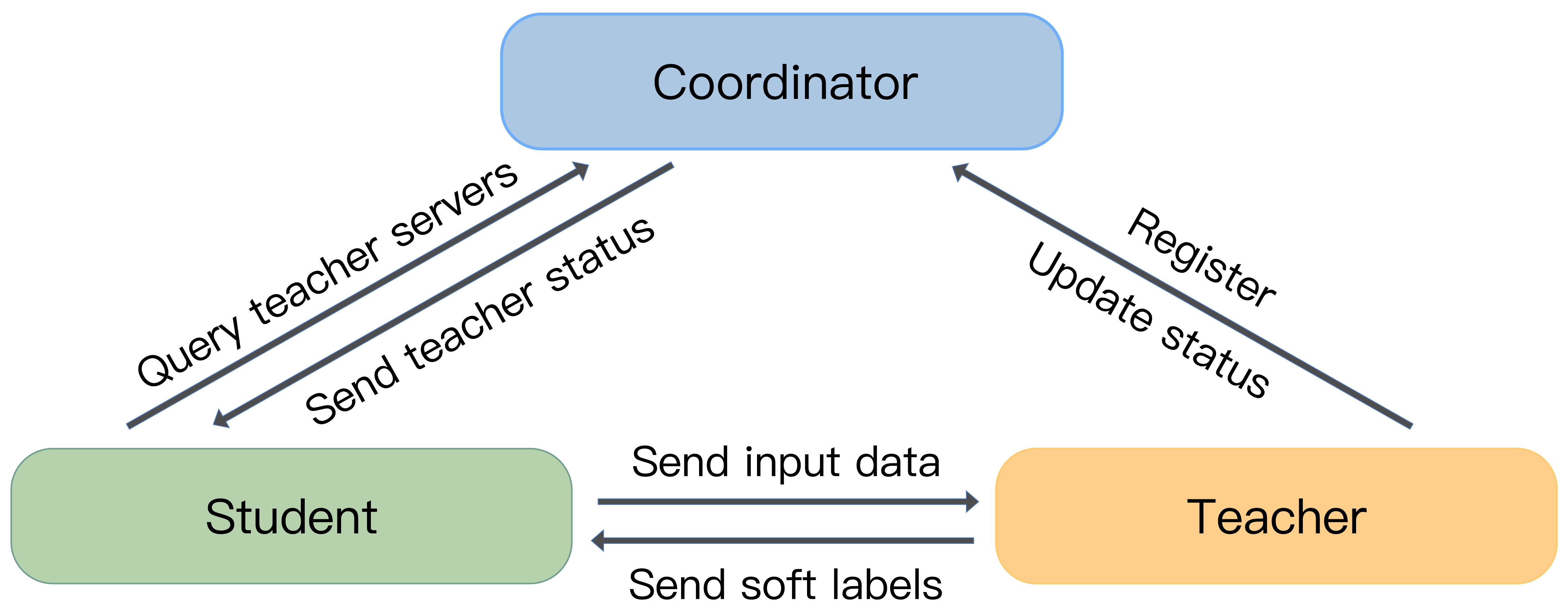}
    \caption{Functional architecture.}
    \label{fig:archi}
\end{figure*}

The computing resources can vary dynamically during the training process \cite{Wu2019}. 
As the computing resources are shared for different workload, some computing resources can be dedicated to a user while some other computing resources are available only when there are no other high-priority workloads.
Some GPU cards may become unavailable because of exceptions or may be withdrawn for other high-priority workloads. 
During the training process of knowledge distillation, some other computing resources may become available to be used when the high-priority workloads are finished.
Thus, it is critical to scale out and to ensure fault-tolerance while exploiting the dynamic computing resources. 
In addition, as diverse types of computing resources are available for the training and the inference process, it is also important to efficiently exploit heterogeneous computing resources for knowledge distillation.
EDL is proposed to train a deep learning model using multiple GPU cards in \cite{Wu2019}, while it does not support the training process of knowledge distillation.

\section{EDL-Dist}
\label{sec:edl}

In this section, we explain our proposed EDL-Dist framework. We first introduce the distributed architecture. Then, we present scheduling and distributed training algorithms for knowledge distillation. Afterward, we detail the fault-tolerance solution to address the exceptions during the training process of knowledge distillation.

\subsection{Architecture}
\label{subsec:farchi}

The architecture of EDL-Dist consists of three modules, i.e., Student, Teacher and Coordinator, as shown in Figure \ref{fig:archi}.
Student contains dedicated computing resources, which are used to train a student model with a distributed or decentralized method.
Teacher is composed of dynamic computing resources, which can be utilized for the inference of the teacher model.
Coordinator manages the distributed training in Student and the data transfer between Student and Teacher. 

A decentralized training algorithm, i.e., ring allReduce \cite{ring-allreduce}, is exploited to realize the parallel training in Student. 
While it takes much time to transfer data across multiple computing resources, the training data is partitioned and cached in the host memory of each server for fast data access. 
Within the distributed training process of Student, only the gradients are transferred while the raw data is kept within each server.
The input data of the student model is composed of three parts, i.e., the input training data, soft labels, and the hard labels. 

The input data is managed by a separated module, i.e., DistilReader, which caches the data in the host memory of computing resources of the Student module in order to accelerate the processing.
DistilReader is an interface among the three modules while it is deployed in each server of the Student module.
DistilReader sends the input data to \liu{Teacher} and receives the soft labels from Teacher (see details in Figure \ref{fig:DistilReader}).
DistilReader retrieves the server information from the Coordinator model to know which Teacher server is connected to which Student server and the availability of the Teacher servers.
In addition, DistilReader searches for other available Teacher servers from Coordinator to replace the unavailable server when a Teacher server becomes unavailable.

The Teacher module contains multiple dynamic computing resources. 
Each resource can become unavailable at any moment because of exceptions or the high priority workload. 
A new server is first registered in the Coordinator module and \liu{then} added in the Teacher module. 
Afterward, a teacher model instance is deployed in the added Teacher server, which can perform the inference process in order to generate corresponding soft labels by processing input training data. 
All the Teacher servers send heartbeat messages to the Coordinator module in order to maintain its alive status until the knowledge distillation is finished.

\liu{
\begin{figure*}[htbp]
    \centering
    \includegraphics[width=0.55\textwidth]{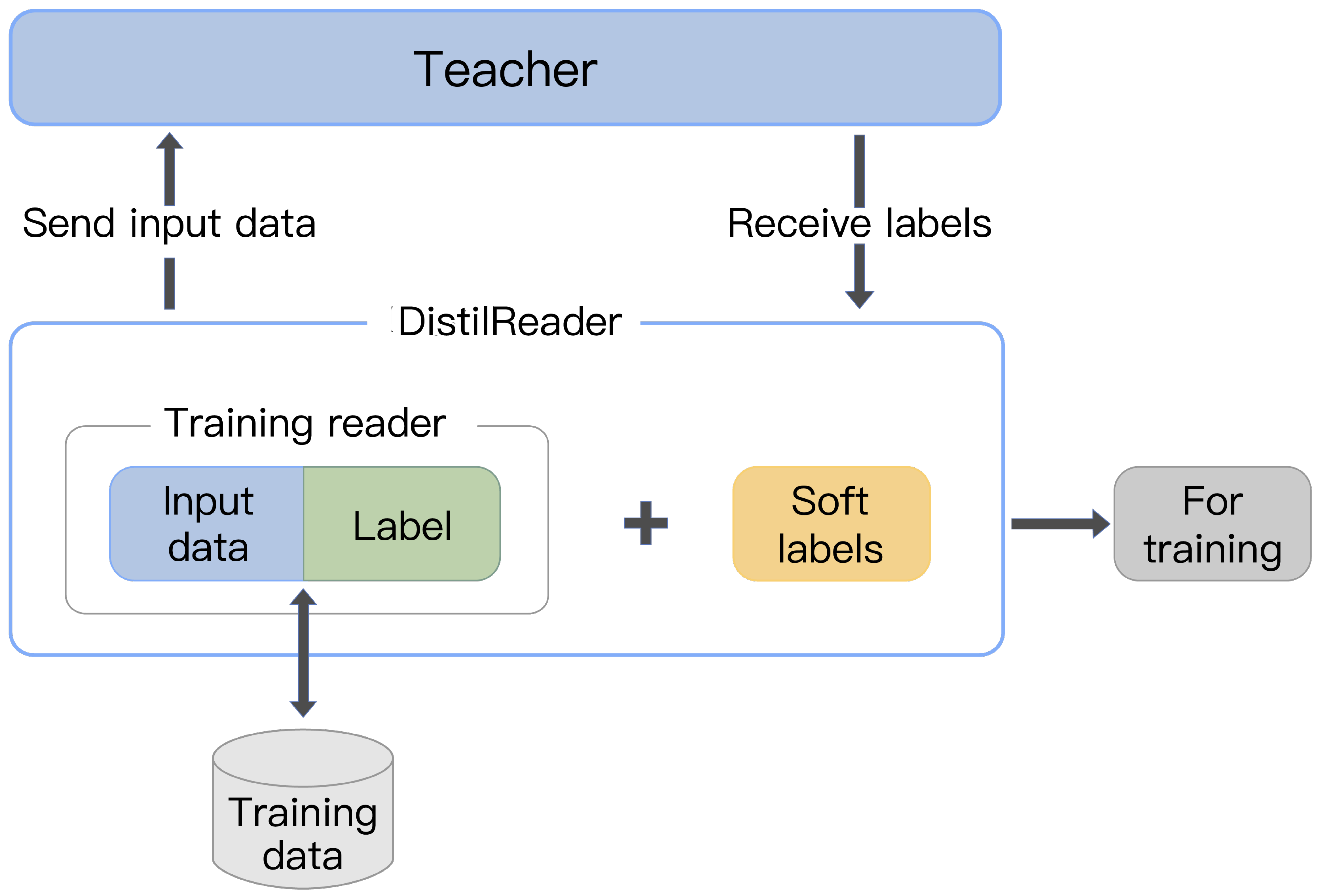}
    \caption{DistilReader service.}
    \label{fig:DistilReader}
\end{figure*}
}

The Coordinator module is composed of two components, i.e., a service manager and a database. 
The database can be an in-memory database in order to achieve efficient data query. 
The service manager receives requests from DistilReaders in the Student module and queries the in-memory database to search for available computing resources in the Teacher module. 
The register information of Teacher servers is directly stored in the in-memory database, while the alive status has a limit, i.e., Time to live (TTL). 
In order to prolong the alive status, each Teacher server continues sending heartbeat messages to the service manager, which updates the data in the in-memory database.
When the Teacher server stops sending heartbeat messages for a long time, which exceeds its TTL, the corresponding status becomes unavailable. 

\subsection{Hybrid Scheduling Algorithm}

It is of much importance to schedule the workloads of the inference process, which generates soft labels by process the input training data, to computing resources in the Teacher module. 
A computing resource is a computing unit to perform the inference or the training, e.g., a CPU core or a GPU card. 
The resource scheduling problem is a typical NP-hard problem \cite{Liu2020}. 
When a Student resource is scheduled to insufficient Teacher resources, the throughput of training process of the student model is limited by the throughput of the inference process in the scheduled Teacher resources.
Otherwise, when a resource in the Student module is scheduled to a much bigger number of Teacher resources, more and more soft labels and corresponding input data are accumulated in the host memory of Student servers, which may occupy large amounts of memory and block the training process while making the system unstable. 
Thus, it is critical to schedule appropriate computing resources in the Teacher module to each Student resource.

In order to ensure the stability of the system, we denote the data size of soft labels in the student module by $S(t)$ and the number of computing resources in the teacher module by $I(t)$, where $t$ represents the time. Then, we have the constraint defined in Formulas \ref{eq:stable} in order to ensure the stability of the system. 

\begin{equation}
\Bar{D} \triangleq \lim_{T \rightarrow \infty} \frac{1}{T}\sum_{t=0}^{T-1}\left(\E\{S(t)\} + \E\{I(t)\}\right) < \infty.
\label{eq:stable}
\end{equation}

In addition, the objective of our scheduling is to minimize the overall data size and the number of the computing resources while ensuring the throughput. In order to ensure the throughput, the data size $S(t)$ should be bigger than a threshold defined by the user and the number of computing resources should be enough in order to produce enough soft labels for the training process of the student module. When the scheduling action is to send input data to Teacher, the data size of soft labels and corresponding input data augment with speed $v$. When the scheduling action is to schedule an additional available Teacher resource to the Student server, the speed is augmented, i.e., $v = v + v'$, where $v'$ can be measured using the previous execution information. When the scheduling action is to stop sending input data to Teacher servers, $v$ becomes $0$. 

\begin{algorithm}[htb] 
\caption{Hybrid Scheduling Algorithm}
\label{alg:hybridScheduling} 
\begin{algorithmic}[1]
\Require number of Teacher resources $n$
\Require lower threshold of the volume of soft labels $lt$
\Require upper threshold of the volume of soft labels $ut$
\State schedule $n$ Teacher resources to the Student resource
\While{knowledge distillation is not terminated}
\State volume $=$ get$\_$volume$($unused soft labels$)$
\If{volume $>$ $ut$}
\State stop sending input data to Teacher servers
\EndIf
\If{volume $==$ 0}
\State schedule an additional available Teacher resource to the Student resource
\EndIf
\If{volume $<$ $lt$}
\State  continue sending input data to Teacher resources
\EndIf
\EndWhile
\end{algorithmic}
\end{algorithm}

We propose
a hybrid scheduling algorithm (see Algorithm \ref{alg:hybridScheduling}), i.e., which combines static and dynamic scheduling methods. We assume historical information on the execution of the training and inference processes. For instance, the throughput of the training in a Student server, e.g., one GPU card in Student, is $t_s$ and the throughput of the inference in a Teacher server, e.g., one GPU card in Teacher, is $t_t$. The throughput gives the number of images or the amount of input data that can be processed in the same resource per time unit by the student model (or the teacher model) without restriction of another module. We assume that the resources in the same module, e.g., Student or Teacher, are of the same type while the types of GPU cards in different modules can be different. We set the number of
Teacher resources as $n = \frac{t_t}{t_s}$ for each Student resource, i.e., we schedule $\lceil n \rceil$ Teacher resources to each Student resource. 
During the training of knowledge distillation, when a Student resource searches for Teacher resources, it is scheduled $\lceil n \rceil$ Teacher resources (Line 1). As the execution environment may vary during the training of knowledge distillation, we dynamically adjust the scheduling (Lines 3-12). We use a
monitoring task in each Student resource to monitor the number of combinations of soft labels and input data (Line 3). 
The occupied volume is calculated based on the number and average size of a combination of input data and soft labels, which can be measured with an offline method. When the growing volume exceeds a predefined upper threshold value (Line 4), the Student resource stops sending input data to the Teacher resource (Line 5) in order to consume the unused soft labels, which can meet the constraint defined in Formula \ref{eq:stable} so as to ensure the system stability.
When the volume decreases to a smaller value than another lower bound threshold value, \liu{the Student} resource continues sending input data to Teacher resources (Lines 10-12). 
The upper threshold and the lower threshold can be set by the user based on the size of storage in the resource of Student.
This mechanism ensures that the number of soft labels remains reasonable in each Student resource, which does not slow down the training or incur memory leaks in the Student resource.
Otherwise, if the resources in Student stay idle in order to wait for the soft labels from Teacher, more Teacher resources are required by the Student resource in order to accelerate the inference in the teacher model (Lines 7-9). When there are available Teacher resources, they are scheduled to the Student resource. 

\subsection{EDL-Dist Algorithm}

We now present our EDL-Dist Algorithm
\ref{alg:EDL-Dist} for the parallel training in each Student resource during the training of knowledge distillation. The input data and the hard label $y$ are retrieved from the host memory (Line 3), which can be done by DistilReader. Then, the soft labels are prepared by the DistReader service from Teacher in Line 4. Based on the hard label and the soft labels, the student model $\theta$ is updated in Line 5. The loss function in each server is a weighted function based on the loss function of the hard labels and the soft labels.
$\lambda$ is the learning rate, which can be set corresponding to the student model. 
Then, an average student model is calculated 
in Line 7. 

\subsection{Fault-tolerance}

\begin{algorithm}[htb] 
\caption{EDL-Dist Algorithm}
\label{alg:EDL-Dist} 
\begin{algorithmic}[1]
\Require hard loss function $\phi$(hard label, hard prediction)
\Require soft loss function $\psi$(soft labels, soft predictions)
\Require hard prediction function $F(\theta, input)$ 
\Require soft prediction function $F'(\theta, input)$ 
\Require learning rate $\eta$ 
\Require weight for hard loss function $\alpha$
\Require weight for soft loss function $\beta$
\Require number of Student resources $N$
\While{not converged}
\For{$\theta_i$ in resource $i$}
\State y, input $=$ get$\_$training$\_$sample()
\State soft$\_$labels $=$ get$\_$soft$\_$labels(input) 
\State $\theta_i = \theta_i - \eta \bigtriangledown_{\theta_i}\{\alpha \phi(y, F(\theta_i, input)) + \beta \psi(soft\_labels, F'(\theta_i, input)\}$  
\EndFor
\State $\theta = \frac{\sum_{j = 1}^{N} \theta_j}{N}$ 
\EndWhile
\end{algorithmic}
\end{algorithm}

We consider the fault-tolerance in Student and Teacher, assuming that Coordinator is always available. 
If the
Coordinator server
is not stable, fault-tolerance can be simply achieved by having multiple instances of the in-memory database deployed in multiple servers using existing frameworks, e.g., Zookeeper \cite{hunt2010zookeeper}. 
If a Teacher resource is not available, its status will become unavailable when its TTL expires in the database. The Teacher resource can become unavailable in three cases.
The first case is before the resource is scheduled to a Student resource.
In this case, EDL-Dist simply ignores this Teacher resource.
The second case is when the Teacher resource is scheduled to a Student resource that does not send input data to it or does not wait for soft labels from it.
In this case, the Student resource will search for another available Teacher resource that is not scheduled to any Student resource.
The third case is when the Teacher resource is scheduled to a Student resource that sends input data to it and is waiting for soft labels from it.
In this case, as presented in Section \ref{subsec:farchi}, the Student resource will search for another available Teacher resource. Once a Teacher resource is re-scheduled to it, the Student resource sends the input data to the Teacher resource again.
When a new Teacher resource is available in Teacher, it is scheduled to a Student resource that is searching for Teacher resources. 
If there is no such Student resource, the Teacher resource will wait for such a Student resource. 

To address fault-tolerance in Student, we exploit
a fail-over mechanism \cite{ozsu2020principles} that uses check-points
during the training of knowledge distillation. A checkpoint is a copy of the student model. Before the training process, a server is selected as a master node and saves the checkpoint at every certain iterations. The checkpoint is saved in a distributed file system, which is accessible to all the Student servers. Each Student server updates the student model in each iteration. Then, when a Student server becomes unavailable or a new Student server is added to Student, the training in all the Student servers stops. Afterward, each Student server loads the student model from the checkpoint and continues the training process. Thus, the consistency of the student model is ensured while addressing fault-tolerance.

\section{Experimental Validation}
\label{sec:exp}

In this section, we present our experimental validation of EDL-Dist in comparison with
online knowledge distillation (Online) (baseline) and normal training (N-training).
We present the experimental setup and then give the results.

\subsection{Experimental Setup}
\label{subsec:implem}

EDL-Dist is implemented based on 
the PaddlePaddle framework \cite{Ma2019} and publicly available at Github \footnote{\url{https://github.com/elasticdeeplearning/edl}}. Student is based on Paddle FleetX \footnote{Paddle Fleet: \url{https://github.com/PaddlePaddle/FleetX}}, which implements the ring allReduce algorithm using NCCL \footnote{NCCL: \url{https://developer.nvidia.com/nccl}} for decentralized training. 
We use Redis \footnote{Redis: \url{https://redis.io/}} as the in-memory database in Coordinator \cite{Liu2019}. 

\liu{
\begin{table}[ht]
\centering
\caption{Summary of experiments. ``*'' represents multiple and that the number of GPU cards is not determined. ``Dmodel'' represents diverse models.
}
\label{tal:summary}
\begin{tabular}{|c|c|c|c|c|c|c|}
\hline
\multirow{2}{*}{Number} & \multirow{2}{*}{Section} & \multicolumn{2}{c|}{Student} & \multicolumn{2}{c|}{Teacher} & \multirow{2}{*}{Objective} \\
\cline{3-6}
 &  & Model & Scale & Model & Scale & \\
\hline
1 & \ref{subsec:heterogeneous} &  MobileNetV3$\_$small & 1 P4 / CPU & Resnet50 & 1 P4 / CPU & Heterogeneous environments \\
\hline
2 &  \ref{subsec:finetuning} & ResNet50 & 1 V100 & ResNet101 & * P4  & Number of Teacher GPU cards \\
\hline
3 & \ref{subsec:comsingleMultiple} & ResNet50 & 8 V100 & ResNet101 & * P4  & Multiple Student GPU cards \\
\hline
4 & \ref{subsec:multipleModels} & Dmodel & 32 V100 & Dmodel & * P4/ * K1200  & Large-scale experimentation \\
\hline
\end{tabular}
\end{table}
}

\begin{table}[ht]
\centering
\caption{\liu{Throughput for different approaches with diverse numbers of CPU cores in Student.}}
\label{tal:hete}
\begin{tabular}{|c|c|c|c|c|}
\hline
CPUcores & N-training & Online & EDL-Dist & Advantage \\
\hline
1 & 14.16 & 5.92 & \textbf{14.34} & \textbf{142.23\%} \\
\hline
2 &  28.44 & 11.51 & \textbf{28.07} & \textbf{143.87\%} \\
\hline
4 & 55.17 & 21.76 & \textbf{54.92} & \textbf{152.39\%} \\
\hline
8 & 101.59 & 37.87 & \textbf{102.40} & \textbf{170.40\%} \\
\hline
16 & 168.42 & 59.94 & \textbf{168.42} & \textbf{180.98\%} \\
\hline
\end{tabular}
\end{table}

We carry out four experiments to show the advantages of EDL-Dist compared with Online and N-training. Online deploys the teacher and student models in the same GPU server. N-training represents the training with GPU cards without knowledge distillation.
In all experiments, we use real datasets, i.e., ImageNet data set \cite{Deng2009}. In the first experiment (Section \ref{subsec:heterogeneous}), we combine CPUs and GPU cards, in order to show that EDL-Dist can efficiently exploit heterogeneous computing resources. 
In the next two experiments, we use ResNet101 \cite{He2016} as the teacher model, ResNet50 \cite{He2016} as the student model, and set the batch size as 32.
The second experiment (Section \ref{subsec:finetuning}) figures out the fine-tuned number of Teacher GPU cards (NVIDIA Tesla P4 GPU card) for each Student GPU card (NVIDIA Tesla V100 GPU card). The single-precision performance, which represents the speed to perform calculation, of P4 is 5.5 Teraflops while that of V100 is 14 Teraflops. 
The third experiment (Section \ref{subsec:comsingleMultiple}) is performed
with 8 V100 GPU cards in Student and various numbers of P4 GPU cards in Teacher for EDL-Dist. 
\liu{In the last experiment (Section \ref{subsec:multipleModels}), we exploit large-scale distributed computing environment to compare EDL-Dist with Online and N-training in terms of throughput and training time with diverse student and teacher models. Large-scale knowledge distillation refers to the support of our approach in with numerous distributed computing resources, i.e. 32 V100 GPU, up to 256 P4, and up to 614 K1200 GPU cards. The experiments are summarized in Table \ref{tal:summary}. }

\subsection{Comparison with Heterogeneous Resources}
\label{subsec:heterogeneous}

To validate that our solution is efficient with heterogeneous computing resources, we experiment with the combination of CPU and GPU cards for knowledge distillation. We take MobileNetV3$\_$small \cite{Koonce2021} as the student model and Resnet50 \cite{He2016} as the teacher model. We use Intel(R) Xeon(R) Gold 6148 CPU @ 2.40GHz CPU cores and a P4 GPU card. We set the batch size as 64 in Student. 

First, we take the P4 GPU card in Teacher and different numbers of CPU cores in Student. 
The results are shown in Table \ref{tal:hete}. 
The throughput of our proposed approach, i.e., EDL-Dist, is similar to that of N-training and significantly outperforms Online (up to 181\%). 
Then, we take the P4 GPU card as the Student GPU card and different numbers of CPU cores as the Teacher resources. 
As shown in Table \ref{tal:hete_normal}, the throughput of EDL-Dist is smaller than that of N-training and Online when Teacher resources are not enough (8). 
The throughput of EDL-Dist is similar to that of N-training and significantly outperforms Online (up to 25.22\%) when the Teacher resources are enough (12 and 16).

\begin{table}[ht]
\centering
\caption{\liu{Throughput for different approaches with diverse numbers of CPU cores in Teacher.}
}
\label{tal:hete_normal}
\begin{tabular}{|c|c|c|c|c|}
\hline
CPUcores & N-training & Online & EDL-Dist & Advantage \\
\hline
8 & 57.14 & 46.04 & 35.68 & -22.50\% \\
\hline
12 &  57.14 & 46.04 & \textbf{52.46} & \textbf{13.94\%} \\
\hline
16 & 55.17 & 46.04 & \textbf{57.65} & \textbf{25.22\%} \\
\hline
\end{tabular}
\end{table}

\begin{figure}[ht]
     \centering
     \begin{subfigure}[b]{0.49\textwidth}
         \centering
         \includegraphics[width=0.9\textwidth]{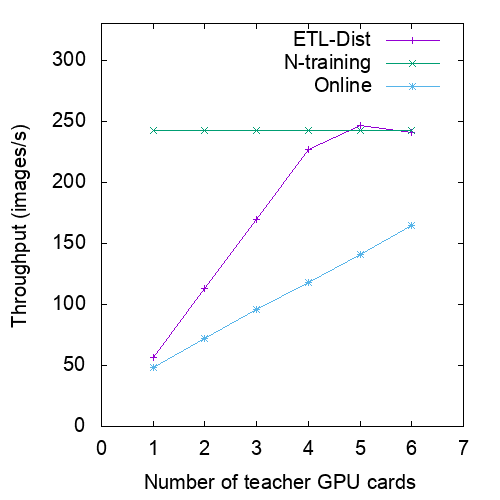}
         \caption{Throughput}
         \label{fig:throughput}
     \end{subfigure}
     \begin{subfigure}[b]{0.49\textwidth}
         \centering
         \includegraphics[width=0.9\textwidth]{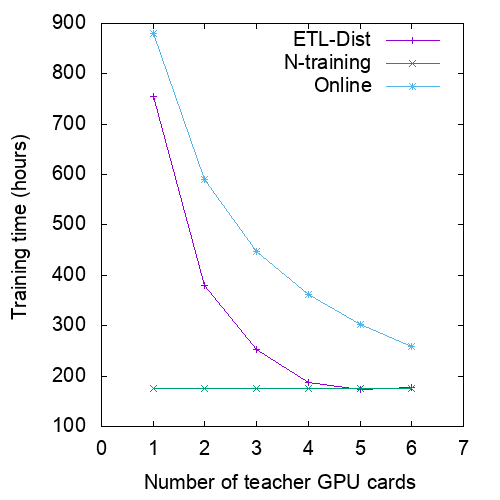}
         \caption{Training time }
         \label{fig:elasticTime}
     \end{subfigure}
     \caption{Fine-tuning with various numbers of P4 Teacher GPU cards.}
\end{figure}

\subsection{Fine-tuning of EDL-Dist}
\label{subsec:finetuning}

The throughput of EDL-Dist increases with the number of Teacher GPU cards. With enough Teacher GPU cards, the throughput of EDL-Dist can be similar to that of N-training. As we add more Teacher GPU cards, the throughput of EDL-Dist becomes a little bit lower as it takes some time to manage unused intermediate soft labels from Teacher.
In order to validate this property of EDL-Dist,
we also experiment using a V100 GPU card in Student and various numbers of P4 GPU cards as Teacher resources. 
The throughput of EDL-Dist is shown in Figure \ref{fig:throughput} when using different numbers of P4 GPU cards. The training time is shown in Figure \ref{fig:elasticTime}. 
Figures \ref{fig:throughput} and \ref{fig:elasticTime} indicate that the fine-tuned number of Teacher resources (P4 GPU card) is 5 when we use a single V100 GPU card as the Student resource.
When the number of Teacher GPU cards is smaller than 5, the throughput increases linearly as number of P4 GPU cards increases, which shows the good scalability of EDL-Dist.
When the number of Teacher GPU cards is greater than 5, the throughput slightly decreases as it takes time to manage unused soft labels in the Student server.
Furthermore, we find that the throughput of Online is much smaller (up to 93.0\%) than that of EDL-Dist and the training time of the Online is much longer (up to 92.9\%) than that of EDL-Dist when the number of Teacher GPU cards is smaller than 8.

\subsection{Comparison with multiple Student GPU Cards}
\label{subsec:comsingleMultiple}

In this experiment, we take 8 V100 GPU cards and 40 - 56 P4 GPU cards for different approaches. We compare the throughput between EDL-Dist, Online and N-training. We take 8 NVIDIA Tesla V100 GPU cards as dedicated Student GPU cards while using 48 P4 NVIDIA Tesla GPU cards as Teacher GPU cards as we find 48 is the appropriate number of Teacher GPU cards as shown in Table \ref{tal:result}. Please note that some GPU servers are dynamically added and removed during the execution of the experiments.

Table \ref{tal:result} shows that the accuracy (1 and 5) of EDL-Dist is similar to that of N-training and Online. Accuracy 1 represents the accuracy of the predicted class with the highest probability. Accuracy 5 represents the accuracy of the top 5 ranked classes based on the probability. The accuracy of EDL-Dist can be slightly higher than that of N-training (Accuracy 5).
\begin{table*}[ht]
\centering
\caption{Experimental Results (accuracy). Accuracy 1 is the accuracy of the predicted class with the highest probability. Accuracy 5 is the accuracy of the top 5 ranked classes based on the probability. }
\label{tal:result}
\begin{tabular}{|c|c|c|c|c|c|}
\hline
         & N-training & Online & EDL-Dist (40) & EDL-Dist (48) & EDL-Dist (56) \\
\hline
\hline
Accuracy 1 & 77.1 & 79.0 & \textbf{79.0} & \textbf{79.0} & \textbf{79.0} \\
\hline
Accuracy 5 & 93.5 & 94.3 & \textbf{94.5} & \textbf{94.5} & \textbf{94.5} \\ 
\hline
\end{tabular}
\end{table*}
\begin{figure}[ht]
     \centering
     \begin{subfigure}[b]{0.49\textwidth}
         \centering
         \includegraphics[width=0.9\textwidth]{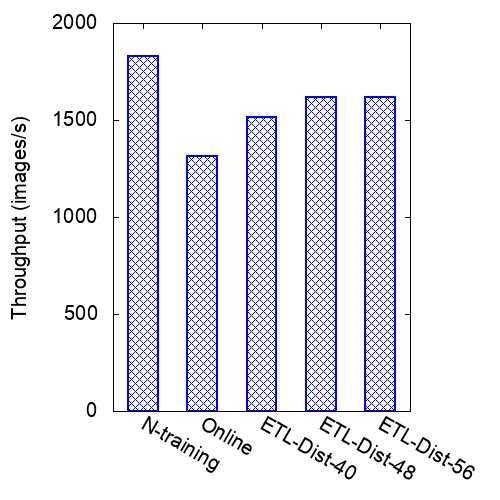}
         \caption{Throughput}
         \label{fig:throughput-m-GPU}
     \end{subfigure}
     \hfill
     \begin{subfigure}[b]{0.49\textwidth}
         \centering
         \includegraphics[width=0.9\textwidth]{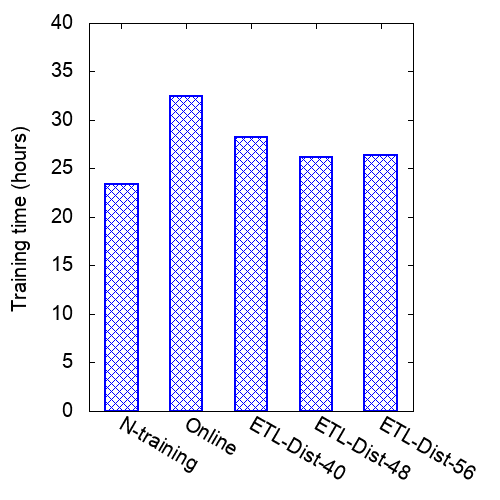}
         \caption{Training time }
         \label{fig:trainingtime}
     \end{subfigure}
     \caption{Experimental results with 8 Student V100 GPU cards and 40(EDL-Dist-40)/48(EDL-Dist-48)/56(EDL-Dist-56) P4 GPU cards.}
\end{figure}
While the student model is trained with the training data and the soft labels with knowledge distillation, the trained student model from knowledge distillation can get more generalization information from the teacher model \cite{Hinton2015}.
Thus, we can efficiently train a student model with higher accuracy (compared with N-training) using EDL-Dist.

In Figure \ref{fig:throughput-m-GPU}, the throughput of EDL-Dist is much higher (23.5\% faster) than that of Online. This shows that EDL-Dist significantly speeds up training compared with the Online while not requiring extra storage resources. The throughput of EDL-Dist is slightly lower than that of N-learning because of some overhead when there are multiple Student GPU cards.
The training time of N-training, Online and EDL-Dist is shown in Figure \ref{fig:trainingtime}. The training time of EDL-Dist (48) is almost the same as that of EDL-Dist (56), which
indicates
that the bottleneck of the number of Teacher GPU cards is 48. With 48 Teacher cards, the training time of EDL-Dist is 19.4\% shorter than that of Online. Compared with N-training, the training time of EDL-Dist is slightly longer (12.8\%). As it takes time to transfer the data from Student servers to multiple Teacher servers, the training time of EDL-Dist is slightly longer than that of N-training.

\begin{figure}[htbp]
    \centering
    \includegraphics[width=0.45\textwidth]{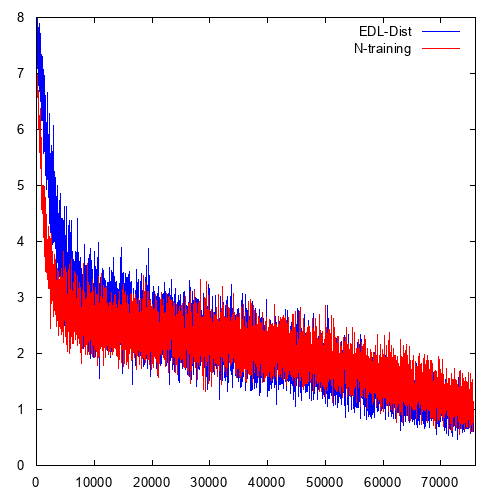}
    \caption{The comparison of training loss for normal training (N-training) and EDL-Dist with 40 (EDL-Dist-40) P4 GPU cards. The training loss of EDL-Dist decreases slightly slower than that of normal training while it reaches almost the same result at the end of 120th epoch.}
    \label{fig:trainingloss}
\end{figure}

We trace the training loss for both EDL-Dist and normal training to understand the convergency as shown in Figure \ref{fig:trainingloss}. We find that the training loss of EDL-Dist decreases lower than that of normal training while it achieves almost the same result at the end of 120th epoch. We believe that the slow decrease at the beginning is caused by the consideration of the soft labels from the teacher model, which contains generalization information. After enough epochs, the training loss of EDL-Dist can be almost the same or slightly smaller than that of normal training.

\subsection{Comparison with Multiple Models}
\label{subsec:multipleModels}

\begin{table}[ht]
\centering
\caption{Throughput for multiple models. ``48'' represents ResNet101$\_$32$\*$48d and ``16'' represents ResNet101$\_$32$\*$16d. 
``Resources'' represents the computing resources, i.e., number and type of GPU cards, in the teacher module. 
The accuracy represents the accuracy of the predicted class with the highest probability. \liu{The time} unit is day. Advantage represents the advantage (times) of EDL-Dist compared with Online.}
\label{tal:multiModels}
\begin{tabular}{|c|c|c|c|c|c|c|c|}
\hline
\multicolumn{2}{|c|}{Model} & \multirow{2}{*}{Resources} & \multicolumn{2}{|c|}{Accuracy} & \multicolumn{2}{|c|}{Execution time} & \multirow{2}{*}{Advantage} \\
\cline{1-2}\cline{4-7}
Teacher & Student & & N-Training & EDL-Dist & Online & EDL-Dist & \\
\hline
48 & Resnet200 & 256 P4 & 80.97\% & 85.13\% & 45 & 14.4 & 3.125 \\
\hline
48 &  HRNet$\_$W64 & 160 P4 & 80.29\% & 85.13\% & 40.9 & 21.4 & 1.911 \\
\hline
16 & Res2Net50 & 614 K1200 & 79.8\% & 83.13\% & 10.1 & 5.25 & 1.923 \\
\hline
16 & HRNet$\_$W48 & 455 K1200 & 78.95\% & 83.64\% & 17.6 & 10.02 & 1.725 \\
\hline
16 & HRNet$\_$W18 & 600 K1200 & 76.92\% & 79.82\% & 12 & 6.50 & 1.846 \\
\hline
16 & Resnet34 & 600 K1200 & 75.98\% & 79.72\% & 16 & 8.9 & 1.798 \\
\hline
16 & Res2Net101 & 614 K1200 & 80.6\% & 83.82\% & 13.5 & 5.93 & 2.275 \\
\hline
\end{tabular}
\end{table}

In order to analyze the throughput for diverse networks, we carried out experimentations with multiple student models, teacher models, and computing resources. In the experimentations, \liu{we} take 32 V100 GPU cards in the student module and various numbers or types of GPU cards in the teacher module. We exploit P4 and K1200\footnote{Parameters for K1200: \url{https://www.nvidia.com/content/dam/en-zz/Solutions/design-visualization/quadro-product-literature/11306_DS_NV_Quadro_K1200_FEB15_NV_US_HR.pdf}} GPU cards in the teacher model. The results are shown in Table \ref{tal:multiModels}. 

From the table, we can see that the throughput of EDL-Dist is always bigger than that of the online distillation while the accuracy of EDL-Dist is slightly higher than that of normal training. The minimum advantage of EDL-Dist is 1.798 times while the maximum advantage of EDL-Dist is 3.125 times in terms of throughput. In addition, the advantage of accuracy of EDL-Dist can be up to 4.85\%, compared with that of normal training. In fact, in the production environment, the ability to use the heterogeneous computing resources can significantly reduce the monetary cost while achieving similar or even better performance, which is of much importance for the usage of EDL-Dist.

\section{Discussion and Future Work}
\label{sec:discussion}

\liu{
In this section, we discuss the following future directions of our current approach:
\begin{itemize}
    \item Synchronization of Ring all-reduce
    \item Non-negligible overhead of EDL-Dist
    \item Distributed computing resources in multiple data centers.
\end{itemize}
}

While the GPU cards become more and more heterogeneous, various types of GPU cards are available for the training process. It is critical to take full advantage of heterogeneous GPU resources for the training process and the inference process of knowledge distillation. In EDL-Dist, we decouple the training process and the inference process, which could exploit diverse types of GPU cards for the training and inference process separately. However, when using heterogeneous GPU resources for the training process, the throughput can be bottlenecked by the slowest GPU cards. 
For instance, the throughput corresponding to the combination of a V100 GPU card and a P4 GPU card is 44.6\% smaller than that of a single V100 GPU for normal training. 
The throughput corresponding to the combination of a V100 GPU card and a P4 GPU card in the student module and a P4 GPU card in the teacher module
is 49\% smaller than using two P4 GPU cards in the teacher module and a single V100 GPU card in the student module for EDL-Dist. We believe this is incurred by the synchronization of Ring all-reduce \cite{ring-allreduce}. Thus, it is critical to enable high efficient decentralized or distributed training process with heterogeneous GPU cards for knowledge distillation and we leave this as future work.

Although the overhead of EDL-Dist is relatively small compared with the gain it brings, it is still non-negligible. As shown in Table  \ref{subsec:comsingleMultiple}, the throughput of the EDL-Dist is 12.8\% smaller than that of normal training, which indicates \liu{that we} can further reduce the overhead of EDL-Dist. As the input data is sent by a student server to its scheduled teacher servers, which can occupies the bandwidth. As a result, the decentralized training process is slowed down. Thus, we can directly put the input data in the teacher module so that the student server only needs to send a reference of the input data so as to reduce the usage of bandwidth for transferring input data. We also leave this as future work.

When we carry out the experiments, we find that not all the GPU resources stay available at a single data center. The current design of EDL-Dist does not consider that the GPU cards are distributed at different data centers. In addition, the data may also be distributed at different data centers or organizations and we can exploit federated learning \cite{Yang2019,liu2021distributed,liu2022Efficient,Zhang2022FedDUAP,Li2022FedHiSyn} in this case. However, few work has been done to enable knowledge distillation with distributed data using federated learning techniques. In the future, we plan to extend EDL-Dist to an environment where the GPU resources and the data are distributed at diverse data centers or organizations with the consideration of federated learning techniques.

\section{Conclusion}
\label{sec:con}
In this paper, we proposed EDL-Dist, an elastic deep learning framework for large scale knowledge distillation.
EDL-Dist has a distributed, fault-tolerant architecture that leverages heterogeneous computing resources. 
We did a thorough validation of our solution by implementing an industrial-strenght prototype of EDL-Dist (available at github) and experimenting with real datasets \liu{and large-scale distributed environments (up to 32 V100, 256 P4, and 614 K1200}.
The experimental results show that EDL-Dist can be 3.125 times faster than online training while its accuracy is a little higher than that of normal training.
In the future, we may exploit federated learning \cite{liu2021distributed} to deal with the decentralized data in order to ensure the data security and privacy.

\nocite{*}
\bibliography{WileyNJD-AMA}

\begin{thebibliography}{10}
\providecommand \doibase [0]{http://dx.doi.org/}%

\bibitem{Villegas17}
Villegas R, Yang J, Zou Y, Sohn S, Lin X, Lee H. Learning to Generate Long-term
  Future via Hierarchical Prediction. In: . 70 of {\it Proceedings of Machine
  Learning Research}. PMLR; 2017\string: 3560-3569.

\bibitem{Szegedy2015}
Szegedy C, Liu W, Jia1 Y, et al. Going deeper with convolutions. In: {IEEE}
  Conf. on Computer Vision and Pattern Recognition ({CVPR}). {IEEE} Computer
  Society; 2015\string: 1-9.

\bibitem{Devlin19}
Devlin J, Chang MW, Lee K, Toutanova K. {BERT:} Pre-training of Deep
  Bidirectional Transformers for Language Understanding. In: Conf. of the North
  American Chapter of the Association for Computational Linguistics: Human
  Language Technologies, ({NAACL-HLT}). ; 2019\string: 4171-4186.

\bibitem{sun2021ernie}
Sun Y, Wang S, Feng S, et al. ERNIE 3.0: Large-scale Knowledge Enhanced
  Pre-training for Language Understanding and Generation. {\it arXiv preprint
  arXiv:2107.02137} 2021.

\bibitem{Caruana2015}
Caruana R, Lou Y, Gehrke J, Koch P, Sturm M, Elhadad N. Intelligible Models for
  HealthCare: Predicting Pneumonia Risk and Hospital 30-day Readmission. In:
  {ACM} {SIGKDD} Int. Conf. on Knowledge Discovery and Data Mining. ;
  2015\string: 1721-1730.

\bibitem{Devlin2018}
Devlin J, Chang M, Lee K, Toutanova K. {BERT:} Pre-training of Deep
  Bidirectional Transformers for Language Understanding. {\it CoRR}
  2018\string; abs/1810.04805.

\bibitem{Bucila2006}
Bucila C, Caruana R, Niculescu{-}Mizil A. Model compression. In: {ACM} {SIGKDD}
  Inte. Conf. on Knowledge Discovery and Data Mining ({SIGKDD}). ; 2006\string:
  535-541

\bibitem{Hinton2015}
Hinton G, Vinyals O, Dean J. Distilling the Knowledge in a Neural Network. In:
  NeurIPS Deep Learning and Representation Learning Workshop. ; 2015.

\bibitem{Gou2020}
Gou J, Yu B, Maybank SJ, Tao D. Knowledge Distillation: {A} Survey. {\it CoRR}
  2020\string; abs/2006.05525.

\bibitem{Mirzadeh2019}
Mirzadeh S, Farajtabar M, Li A, Ghasemzadeh H. Improved Knowledge Distillation
  via Teacher Assistant: Bridging the Gap Between Student and Teacher. {\it
  CoRR} 2019\string; abs/1902.03393.

\bibitem{li2021knowledge}
Li X, Xiong H, Chen Z, et al. Knowledge Distillation with Attention for Deep
  Transfer Learning of Convolutional Networks. {\it ACM Transactions on
  Knowledge Discovery from Data ({TKDD})} 2021\string; 16(3)\string: 1-20.

\bibitem{ozsu2020principles}
{\"O}zsu MT, Valduriez P. {\it Principles of distributed database systems}.
\newblock Springer.
\newblock 4~ed. 2020.

\bibitem{liu2020two}
Liu J, Bondiombouy C, Mo L, Valduriez P. Two-phase scheduling for efficient
  vehicle sharing. {\it IEEE Transactions on Intelligent Transportation Systems
  ({TITS})} 2022\string; 23(1)\string: 457-470.

\bibitem{dong2022elastic}
Dong D, Liu J, Wang X, et al. Elastic Deep Learning Using Knowledge
  Distillation with Heterogeneous Computing Resources. In: European Conference
  on Parallel Processing workshop. ; 2022\string: 116-128.

\bibitem{Zmora2019}
Zmora N, Jacob G, Zlotnik L, Elharar B, Novik G. Neural Network Distiller: {A}
  Python Package For {DNN} Compression Research. {\it CoRR} 2019\string;
  abs/1910.12232.

\bibitem{Zhang2018}
Zhang Y, Xiang T, Hospedales TM, Lu H. Deep Mutual Learning. In: IEEE Conf. on
  Computer Vision and Pattern Recognition ({CVPR}). ; 2018\string: 4320-4328.

\bibitem{Chen2020}
Chen D, Mei J, Wang C, Feng Y, Chen C. Online Knowledge Distillation with
  Diverse Peers. In: Conf. on Artificial Intelligence ({AAAI}). ; 2020\string:
  3430-3437.

\bibitem{Cheng2017}
Cheng Y, Wang D, Zhou P, Zhang T. A Survey of Model Compression and
  Acceleration for Deep Neural Networks. {\it CoRR} 2017\string;
  abs/1710.09282.

\bibitem{Anil2018}
Anil R, Pereyra G, Passos A, Orm{\'{a}}ndi R, Dahl GE, Hinton GE. Large scale
  distributed neural network training through online distillation. In: Int.
  Conf. on Learning Representations ({ICLR}). ; 2018.

\bibitem{Madiajagan2019}
Madiajagan M, Raj SS. Chapter 1 - Parallel Computing, Graphics Processing Unit
  (GPU) and New Hardware for Deep Learning in Computational Intelligence
  Research. In:  Sangaiah AK. \kern-2pt, ed. {\it Deep Learning and Parallel
  Computing Environment for Bioengineering Systems}Academic Press.  2019 (pp. 1
  - 15).

\bibitem{Narayanan2019}
Narayanan D, Harlap A, Phanishayee A, et al. PipeDream: Generalized Pipeline
  Parallelism for DNN Training. In: ACM Symposium on Operating Systems
  Principles ({SOSP}). ; 2019\string: 1–15.

\bibitem{li2014communication}
Li M, Andersen DG, Smola AJ, Yu K. Communication efficient distributed machine
  learning with the parameter server. In: Advances in Neural Information
  Processing Systems ({NeurIPS}). ; 2014\string: 19-27.

\bibitem{liu2021heterps}
Liu J, Wu Z, Yu D, et al. Heterps: Distributed deep learning with reinforcement
  learning based scheduling in heterogeneous environments. {\it arXiv preprint
  arXiv:2111.10635} 2021.

\bibitem{Valiant1990}
Valiant LG. A Bridging Model for Parallel Computation. {\it Communications of
  the {ACM}} 1990\string; 33(8)\string: 103-111.

\bibitem{Zhu2020}
Zhu R, Yang S, Pfadler A, Qian Z, Zhou J. Learning Efficient Parameter Server
  Synchronization Policies for Distributed {SGD}. In: Int. Conf. on Learning
  Representations ({ICLR}). ; 2020.

\bibitem{Ho2013}
Ho Q, Cipar J, Cui H, et al. More Effective Distributed {ML} via a Stale
  Synchronous Parallel Parameter Server. In: ; 2013\string: 1223-1231.

\bibitem{Lian2018}
Lian X, Zhang W, Zhang C, Liu J. Asynchronous Decentralized Parallel Stochastic
  Gradient Descent. In: Machine Learning Research. ; 2018\string: 3049-3058.

\bibitem{ring-allreduce}
Gibiansky A. Bringing HPC Techniques to Deep Learning.
  \url{https://andrew.gibiansky.com/blog/machine-learning/baidu-allreduce/};
  2017.
\newblock Accessed: 2020-08-12.

\bibitem{Lian2017}
Lian X, Zhang C, Zhang H, Hsieh C, Zhang W, Liu J. Can Decentralized Algorithms
  Outperform Centralized Algorithms? {A} Case Study for Decentralized Parallel
  Stochastic Gradient Descent. In: Annual Conf. on Neural Information
  Processing Systems ({NeurIPS}). ; 2017\string: 5330-5340.

\bibitem{sergeev2018horovod}
Sergeev A, Balso MD. Horovod: fast and easy distributed deep learning in
  {TensorFlow}. {\it arXiv preprint arXiv:1802.05799} 2018.

\bibitem{Wu2019}
Wu Y, Ma K, Yan X, Liu Z, Cheng J. Elastic deep learning in multi-tenant {GPU}
  cluster. {\it CoRR} 2019\string; abs/1909.11985.

\bibitem{Liu2020}
Liu L, Yu H, Sun G, Luo L, Jin Q, Luo S. Job scheduling for distributed machine
  learning in optical WAN. {\it Future Generation Computer Systems}
  2020\string; 112\string: 549 - 560.

\bibitem{hunt2010zookeeper}
Hunt P, Konar M, Junqueira FP, Reed B. ZooKeeper: Wait-free Coordination for
  Internet-scale Systems. In: USENIX annual technical conf. ; 2010\string: 11.

\bibitem{Ma2019}
Ma Y, Tian~Wu aDY, Wang H. PaddlePaddle: An Open-Source Deep Learning Platform
  from Industrial Practice. {\it Frontiers of Data and Computing} 2019\string;
  1(1)\string: 105.

\bibitem{Liu2019}
Liu J, Pineda{-}Morales L, Pacitti E, et al. Efficient Scheduling of Scientific
  Workflows Using Hot Metadata in a Multisite Cloud. {\it {IEEE} Transactions
  on Knowledge and Data Engineering ({TKDE})} 2019\string; 31(10)\string:
  1940-1953.

\bibitem{Deng2009}
Deng J, Dong W, Socher R, Li L, Li K, Li F. ImageNet: {A} large-scale
  hierarchical image database. In: {IEEE} Conf. on Computer Vision and Pattern
  Recognition {(CVPR}). ; 2009\string: 248-255.

\bibitem{He2016}
He K, Zhang X, Ren S, Sun J. Deep Residual Learning for Image Recognition. In:
  {IEEE} Conf. on Computer Vision and Pattern Recognition ({CVPR}). ;
  2016\string: 770-778.

\bibitem{Koonce2021}
Koonce B. {\it MobileNetV3}\string: 125-144; Apress .
\newblock 2021.

\bibitem{Yang2019}
Yang Q, Liu Y, Chen T, Tong Y. Federated Machine Learning: Concept and
  Applications. {\it {ACM} Transactions on Intelligent Systems and Technology
  ({TIST})} 2019\string; 10(2)\string: 12:1-12:19.

\bibitem{liu2021distributed}
Liu J, Huang J, Zhou Y, et al. From distributed machine learning to federated
  learning: a survey. {\it Knowledge and Information Systems} 2022\string;
  64(4)\string: 885-917.

\bibitem{liu2022Efficient}
Zhou C, Liu J, Jia J, et al. Efficient Device Scheduling with Multi-Job
  Federated Learning. In: AAAI Conf. on Artificial Intelligence ({AAAI}). ;
  2022\string: 1-9.
\newblock To appear.

\bibitem{Zhang2022FedDUAP}
Zhang H, Liu J, Jia J, Zhou Y, Dai H. FedDUAP: Federated Learning with Dynamic
  Update and Adaptive Pruning Using Shared Data on the Server. In: Int. Joint
  Conf. on Artificial Intelligence ({IJCAI}). ; 2022\string: 1-7.
\newblock To appear.

\bibitem{Li2022FedHiSyn}
Li G, Hu Y, Zhang M, et al. FedHiSyn: A Hierarchical Synchronous Federated
  Learning Framework for Resource and Data Heterogeneity. In: Int. Conf. on
  Parallel Processing ({ICPP}). ; 2022\string: 1-10.
\newblock To appear.

\end{thebibliography}

\end{document}